# Untying molecular friction knots.

Serdal Kirmizialtin and Dmitrii E. Makarov*

*Department of Chemistry and Biochemistry and Institute for Theoretical Chemistry,*

*University of Texas at Austin, Austin, Texas, 78712*

**Abstract**

Motivated by recent advances in single molecule manipulation techniques that enabled several groups to tie knots in individual polymer strands and to monitor their dynamics, we have used computer simulations to study "friction knots" joining a pair of polymer strands. The key property of a friction knot splicing two ropes is that it becomes jammed when the ropes are pulled apart. In contrast, molecular friction knots eventually become undone by thermal motion. We show that depending on the knot type and on the polymer structure, a friction knot between polymer strands can be strong (the time $\tau$ the knot stays tied increases with the force $F$ applied to separate the strands) or weak ($\tau$ decreases with increasing $F$). We further present a simple model explaining these behaviors.

Molecular knots tied in individual polymer strands have fascinated researchers from many fields, see, e.g.,.[1-11] Recent progress in single molecule manipulation techniques (reviewed in) [12-14] has enabled several experimentalists to tie a variety knots in single biopolymer strands by using optical tweezers [15, 16]. With these techniques, it is possible to create individual polymeric structures of complex topology and to study their dynamics under mechanical tension. Such structures may prove useful in nanotechnology applications. In addition, knotted DNA structures are common in biology; Studies of the intra-strand interactions in molecular knots may provide new insights into the molecular forces that control the DNA dynamics and the organization of the chromatin fiber[3].

Motivated by the experimental advances, this paper discusses the dynamics of friction knots formed by a pair of polymer molecules. Friction knots, such as the square knot shown in Fig. 1, are commonly used by sailors and climbers to join two ropes together. [Note that they are not "true knots" in the topological sense]. Pulling at the ends of the ropes in Fig. 1 jams the knot so that the ropes remain connected regardless of the applied force. An elegant theory exists[17], which explains this behavior and shows that if the friction coefficient between the ropes exceeds a certain knot dependent critical value then the two ropes will not come apart no matter how hard one pulls on them. This theory also explains why a slight modification of the square knot known as granny knot (also shown in Fig. 1), will be a very poor way of splicing two ropes that will fail at a low force. Here, we would like to find out whether similar behavior could be observed on a microscopic scale, where ropes are replaced by polymer molecules.

A friction knot, scaled down to molecular dimensions, will no longer hold indefinitely under applied tension. Indeed, the knotted conformations of the two molecules shown in Fig. 1 would be thermodynamically unfavorable under an arbitrarily low force $\mathbf{F}$ as the free energy of the system contains the term $-\mathbf{FR}$ ($\mathbf{R}$ being the distance vector between the ends of the strands at which the force is applied, see Fig. 1), which can decrease indefinitely when the two strands are separated. Microscopically, eventual failure of the knot is caused by thermal fluctuations – a macroscopic analog of this would be to pull on the ropes joined by a knot while shaking them vigorously, which would obviously facilitate their separation.

Nevertheless, signatures of the knot jamming effect can be found when examining the *dynamics* of molecular knots. To compare the dynamic response of macroscopic and microscopic knots to tension, note that the strength of macroscopic knots is related to static friction, which impedes relative sliding of the two strands[1, 17]. In contrast, there is no static friction between molecules. Instead the inter-chain "internal friction" is a consequence of the bumpiness of the energy landscape of the interacting polymers[18]. Two intertwined chains may become trapped in conformations corresponding to local energy minima. The sliding of one relative to the other is then accomplished via thermally activated transitions from one local minimum to the next. Unless the temperature is zero such transitions will happen even if the force is arbitrarily small.

However just as the static friction force between two ropes joined by a friction knot increases with the applied tension[17], the barriers to the sliding of one polymer strand with respect to the other may increase. We therefore expect that it may take *longer* to unravel a molecular friction knot when the applied tension is *higher*. We will refer to this as "strong knot" behavior as opposed to "weak knots" that untie faster when higher force is applied. Strong knots are reminiscent of molecular "catch-bonds" observed in forced dissociation of some biomolecular complexes (see, e.g., [19, 20] and refs. therein).

To test our prediction, we have performed computer experiments examining the tension-induced dynamics of various knots tied between two polymer strands. We used a polymer model, in which monomers were represented as single beads. The potential energy of a strand, as a function of the position $\mathbf{r}_i$, $i=1, \ldots, N$, of each bead, is given by:

$$V(\mathbf{r}_1, \mathbf{r}_2, \ldots, \mathbf{r}_N) = V_{bond} + V_{bend} + V_{non\text{-}bonded}$$

The potential $V_{bond}$ accounts for the connectivity of the chain and assumes that each bond is a stiff harmonic spring,

$$V_{bond} = \sum_{i=2}^{N} k_b (|\mathbf{u}_i| - l_{i,i-1})^2 / 2.$$

Here $\mathbf{u}_i = \mathbf{r}_i - \mathbf{r}_{i-1}$ is the bond vector and $l_{i,i-1}$ is the equilibrium bond length given by: $l_{i,i-1} = \rho_i + \rho_{i-1}$, where $\rho_i$, $\rho_{i-1}$ are the effective sizes (i.e., the van der Waals radii) of the $i$-th and ($i$-1)-th monomers. We have constructed polymer chains consisting of two types of beads (see below), bead A and bead B with $\rho_A = \sigma/2$ and $\rho_B = 5\sigma/4$, where $\sigma$ is the equilibrium A-A bond length. The spring constant is taken to be $k_b = 500\ \varepsilon/\sigma^2$, where $\varepsilon$ sets the energy scale. The bending potential is:

$$V_{bend} = \sum_{i=2}^{N-1} k_\theta (\theta_i - \theta_0)^2 / 2$$

where $\theta_0 = \pi$ is the equilibrium bending angle, $\theta_i$ is the angle between $\mathbf{u}_i$ and $\mathbf{u}_{i+1}$, and $k_\theta$ is the bending spring constant. The value $k_\theta = 5\varepsilon/(rad)^2$ used in our simulations corresponds to a persistence length of 15 monomers at temperature T=0.4 $\varepsilon/\sigma$.

The energy $V_{\text{non-bonded}}$ describes the interaction between pairs of monomers that are not covalently bonded. We took this interaction to be purely repulsive:

$$V_{\text{non-bonded}} = \sum_{|i-j|\geq 2} \varepsilon \left[\left(\frac{\rho_i + \rho_j}{|\mathbf{r}_i - \mathbf{r}_j|}\right)^{12}\right].$$

In addition to interactions among non-bonded monomers within each chain, the same pairwise potential was used to describe the interactions between pairs of monomers belonging to different chains.

We further assumed that the dynamics of the chains were governed by the Langevin equation of the form $m\ddot{\mathbf{r}}_i = -\xi\dot{\mathbf{r}}_i - \partial V/\partial \mathbf{r}_i + \mathbf{f}_r(t)$, where $\mathbf{r}_i$ is the position of the $i$-th bead, $m$ is its effective mass, $\xi$ is the friction coefficient, for which we chose the value $\xi = 2.0\left(\sigma^2/m\varepsilon\right)^{-1/2}$, and $\mathbf{f}_r(t)$ is a random δ-correlated force satisfying the fluctuation-dissipation theorem. This equation was solved by using the velocity Verlet algorithm as described in[21]. In reporting our data below, we use dimensionless units of energy, distance, time, and force respectively equal to $\varepsilon$, $\sigma$, $\tau_0 = (m\sigma^2/\varepsilon)^{1/2}$, and $F_0 = \varepsilon/\sigma$.

In the beginning of each simulation, we connect the two strands by a square or granny knot positioned such that the contour length of the polymer chain between the knot and the end of each strand is the same. A force $F_p = 4.0\, F_0$ is then applied to the ends of one strand and $-F_p$ to the ends of the other strand, for an initial time of $t_p = 2000\, \tau_0$. This force pre-tensions the knot without considerably affecting its initial location relative to the ends of each polymer. After preparing the initial state of the knot this way, we start simulation at $t = 0$, with a force $F$ applied to the first bead ($i=1$) of one chain and the opposite force acting on the last bead ($i=N$) of the other one. We monitor the presence of the knot by projecting the polymers' configuration onto a plane that is parallel to the direction of the force and computing the chain intersections in this plane[8]. The knot disappears when the number of intersections falls below 6. This allows us to measure the time $\tau$ before the knot disappears.

We also monitor the distance $R$ between the monomers at which the force is applied. The observed trajectories $R(t)$ typically display an initial transient behavior that has to do with the particular way the knot is prepared followed by an approximately linear increase in the distance $R$. Discarding the transient part, the average strand separation rate, $\langle dR/dt \rangle$, is a convenient way to describe the knot's response to a pulling force. Typical dynamics of the square knot observed in our simulations are shown in Fig. 2 (also see the supplementary video files).

We found that the square knot formed between two identical homopolymer strands, $(A)_{88}$ or $(B)_{88}$, is a weak knot, for the particular polymer model we used. Our interpretation of this observation is that the energy landscape associated with the interaction of two homopolymer strands within our model is not rugged enough to produce the expected jamming effect.

We then achieved a more rugged energy landscape by constructing heteropolymers of the form $AAA(ABAAA)_{17}$. The idea that variable size of monomers can result in a bumpier energy landscape can be intuitively understood by considering the following experiment the reader can perform with any suitable piece of jewelry: Tie a square knot between two strands of beads on a string and then attempt to separate the strands by pulling at their ends. The strands tend to snag in configurations that in fact correspond to local energy minima. This tendency to snag is higher if the beads are of variable size, as compared to equal-size beads.

Figure 3 shows the average time $\tau(F)$ it took for the two polymer strands forming a square knot to become separated in our simulations, as a function of the pulling force. When both strands were homopolymers ($A_{88}$ or $B_{88}$), this time decreased monotonically and was approximately inversely proportional to $F$. However when each strand was a heteropolymer $AAA(ABAAA)_{17}$, the separation time initially decreased and then increased with the increasing force thus exhibiting the strong knot behavior at high forces.

Like its macroscopic counterpart, the molecular version of the granny knot fails much more easily than the square knot: When the same two heteropolymer strands were joined by the granny knot, the time $\tau(F)$ first decreased with the increasing force and then became nearly force-independent, as also shown in Fig. 3.

It is reasonable to expect that the slowdown in the untying dynamics of molecular friction knots would be more pronounced at low temperatures, when there is less thermal motion. Indeed, this is what we see in Fig. 4, which explores the dependence of the mean strand separation time $\tau(F)$ on temperature.

To rationalize the above findings and to understand how forces can influence the knot dynamics, consider the simplest model that relates the effective friction to the features of the energy landscape[18]. Suppose the relative sliding of the two strands can be viewed as one-dimensional diffusive motion along the coordinate $R$; The Brownian dynamics along $R$ is described by the stochastic equation $\eta \dot{R} = F - dV_F(R)/dR + f_r(t)$, where $\eta$ is a friction coefficient and $f_r(t)$ is a random force that satisfies the standard fluctuation-dissipation relationship. The potential $V_F(R)$ is our model for the corrugated energy landscape for inter-strand interaction. We will assume it to be periodic, $V_F(R) = v(F)\sin(2\pi R/a)$. [A random potential may be a better model; however it will not qualitatively change our conclusions]. The effect of the force $F$ is to tilt the overall potential, $V_F(R) \to V_F(R) - FR$, and also to change the degree of corrugation of the inter-strand potential, which is described by the parameter $v(F)$.

The average velocity of diffusion along $R$ can be evaluated exactly[22]:

$$\langle dR/dt \rangle = \frac{k_B T}{\eta a}(1-e^{-aF/k_B T})\left\{\int_{x_0}^{x_0+1} dx \int_{x-1}^{x} dy\, e^{(v(F)\sin(2\pi x)-v(F)\sin(2\pi y)+Fay-Fax)/k_B T}\right\}^{-1}, \quad (1)$$

where the result does not depend on $x_0$. The amplitude $v(F)$ should increase with $F$ to describe the tendency of the potential to become more corrugated. For low enough forces we can assume this to be a linear function: $v(F) = Fd$, where the coupling parameter $d$ has the units of length. Depending on the value of $d$, there are two regimes illustrated in Fig. 5a:

(1). If $d < d_c = a/2\pi$ then the potential $V_F(R) - FR$ is barrierless and decreases monotonically with $F$. In this case the sliding speed $\langle dR/dt \rangle$ should increase with the increasing force and the strand separation time should decrease monotonically. This is the weak knot behavior.

(2) However if $d > a/2\pi$ then the barriers in $V_f(R) - RF$ will become higher when $F$ is increased. When they are higher than $k_B T$ we expect this to lead to a decrease in $\langle dR/dt \rangle$. This is the strong knot regime. At low forces, evaluating Eq. 1 analytically to 1$^{st}$ order in $F$ we see that it approaches the free drift limit $\langle dR/dt \rangle = \langle dR/dt \rangle_{free} = F/\eta$.

The average sliding speed $\langle dR/dt \rangle$ thus first increases and then decreases with $F$, which explains the minimum of $\tau(F)$ seen in Figs. 3-4.

From Eq. 1, inter-strand interaction slows down the strand separation by the factor:

$$\frac{\langle dR/dt \rangle_{free}}{\langle dR/dt \rangle} = \frac{\tilde{F}}{1-e^{-\tilde{F}}} \left\{ \int_{x_0}^{x_0+1} dx \int_{x-1}^{x} dy\, e^{(d/a)\tilde{F}[\sin(2\pi x)-\sin(2\pi y)]+\tilde{F}(y-x)} \right\}, \quad (2)$$

which only depends on two parameters, the dimensionless force $\tilde{F} = Fa/k_B T$ and the dimensionless coupling strength $d/a$. We therefore expect that if we plot the drift velocity (normalized by $\langle dR/dt \rangle_{free}$) vs. $F/T$, the resulting plot will be a universal curve that does not depend on the temperature. As seen from Fig. 5b, this prediction is indeed correct, supporting the validity of the simple one-dimensional model as a description of the square knot dynamics.

Maddocks & Keller theory [17] predicts that the friction coefficient between two ropes must exceed a knot-type dependent critical value for the knot to hold. Our model's prediction for molecular friction knots is very similar: The value of the coupling parameter $d/a$ depends on both the knot type (which determines how the tension in the polymer strands is transmitted into the intra-strand effective friction[17]) and the nature of the polymer strands. As noted above, in order for a knot to be strong, this parameter must exceed a certain critical value. The weakness of the granny knot and of the square knot between two homopolymer strands observed here can be interpreted as a consequence of the coupling being too low.

**Acknowledgments.** We thank Ioan Andricioaei, Oscar Gonzalez, Sergy Grebenshchikov, John Maddocks, and Peter Rossky for helpful discussions. This work was supported by the Robert A. Welch Foundation and by the National Science Foundation CAREER award to DEM. The CPU time was provided by the Texas Advanced Computer Center.

**Figure Captions.**

**Figure 1**. **The square knot and the granny knot**. Although the granny knot is very similar to the square knot, it will fail at a low force while the square knot will only become tighter as the tension in the ropes is increased.

**Figure 2. Dynamics of knot untying.** Snapshots of two polymer strands observed in a Langevin Dynamics simulation. The time increases from top to bottom. The two strands, each with the sequence AAA(ABAAA)$_{17,}$ were initially joined by a square knot and subsequently pulled apart. Two animations of the dynamics of the square and the granny knots observed in simulations are included in Supplementary Information. The snapshots and the movies were generated with the help of the PyMol software[23].

**Figure 3. Effect of polymer sequence and of the knot type on the untying time**. The mean time $\tau$ for the untying of the square and the granny knots as a function of the force pulling the polymer strands apart for different polymer chains and different knots. Since the time is proportional to the contour length $L$ of the polymer, the plotted value of $\tau$ is normalized by $L$. The units $F_0$ and $\tau_0$ are explained in the Methods section.

**Figure 4. Effect of temperature on the knot untying time**. The mean time $\tau$ for the untying of the square knot as a function of the force pulling the polymer strands apart at different temperatures. The units of temperature are explained in the Methods section.

**Figure 5. The tilted periodic potential model**. (a). Salient features of molecular friction knot dynamics can be rationalized by considering the model of Brownian dynamics in a periodic potential tilted by the force F: $Fd \sin(2\pi R/a) - FR$. For sufficiently small values of $d$, the potential is barrierless (cf. the dashed lines corresponding to the case $d=0.1\ a$). However when $d$ is sufficiently large, the potential becomes more bumpy as the force $F$ increases (cf. solid lines corresponding to $d=0.35a$) and, as a result, the overall drift velocity decreases with the increasing $F$.

(b). Interaction between two polymer strands within the square knot slows down their separation by a factor $\langle dR/dt \rangle_{free} / \langle dR/dt \rangle$, which is plotted as a function of $F/T$. According to the tilted periodic potential model, the data plotted this way should form a universal curve that does not depend on the temperature. Indeed, we find this to be the case here.

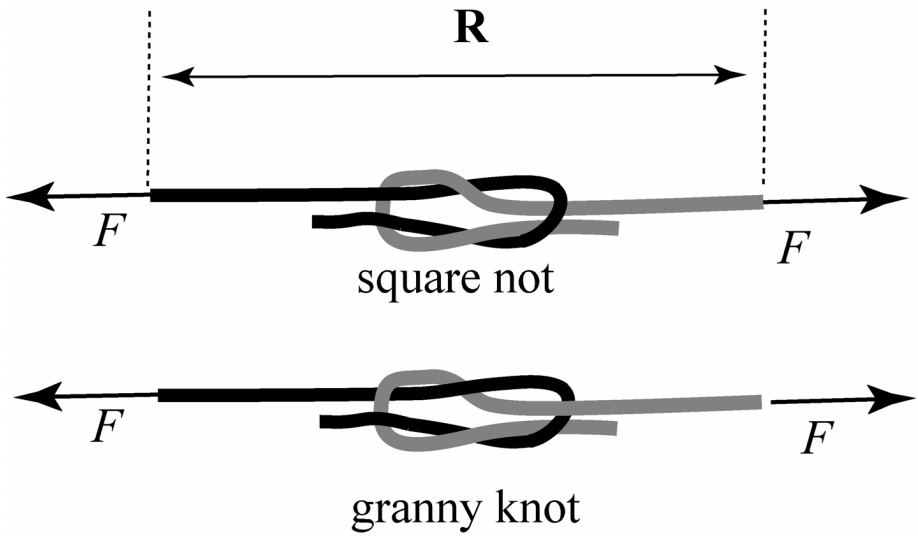

**Figure 1**

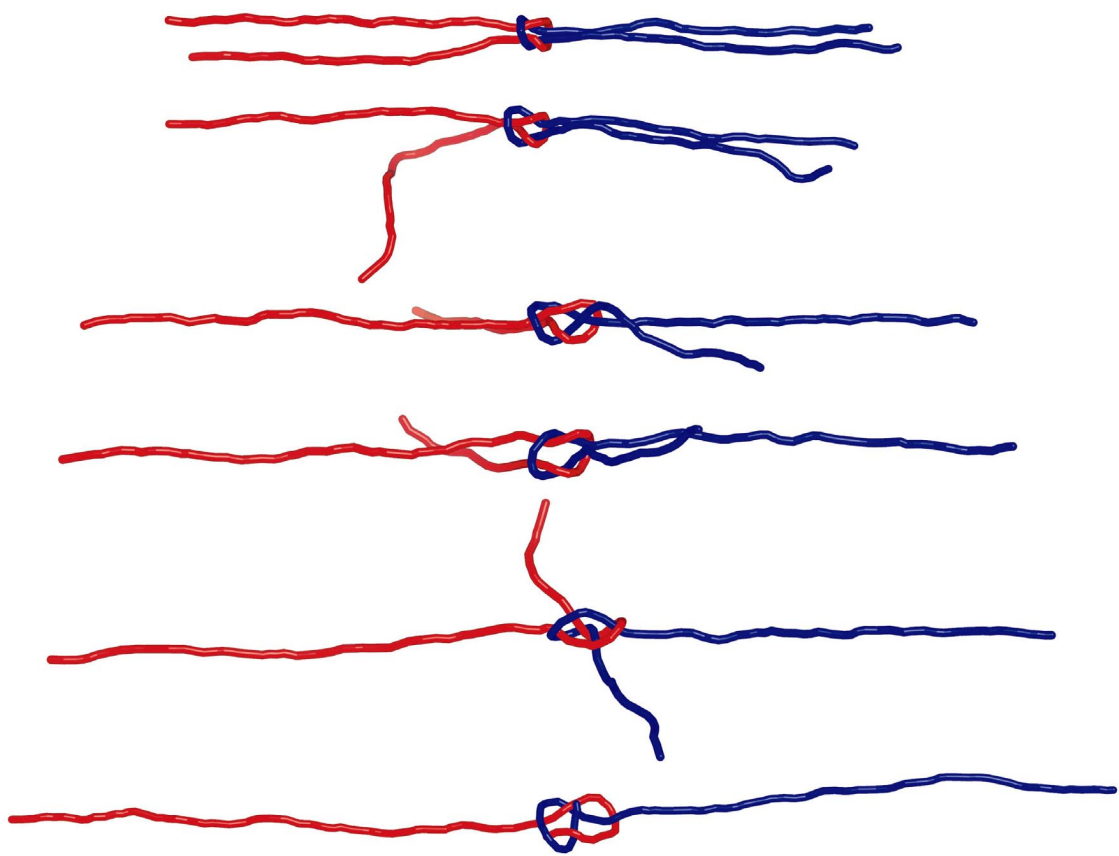

**Figure 2**

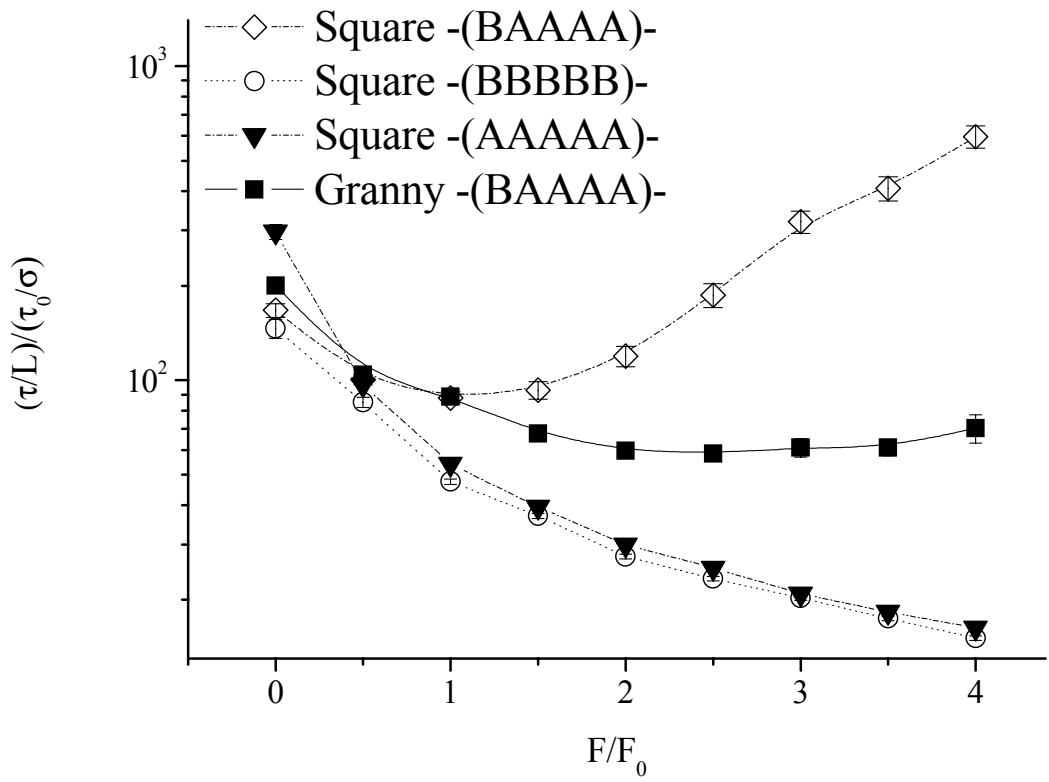

**Figure 3**

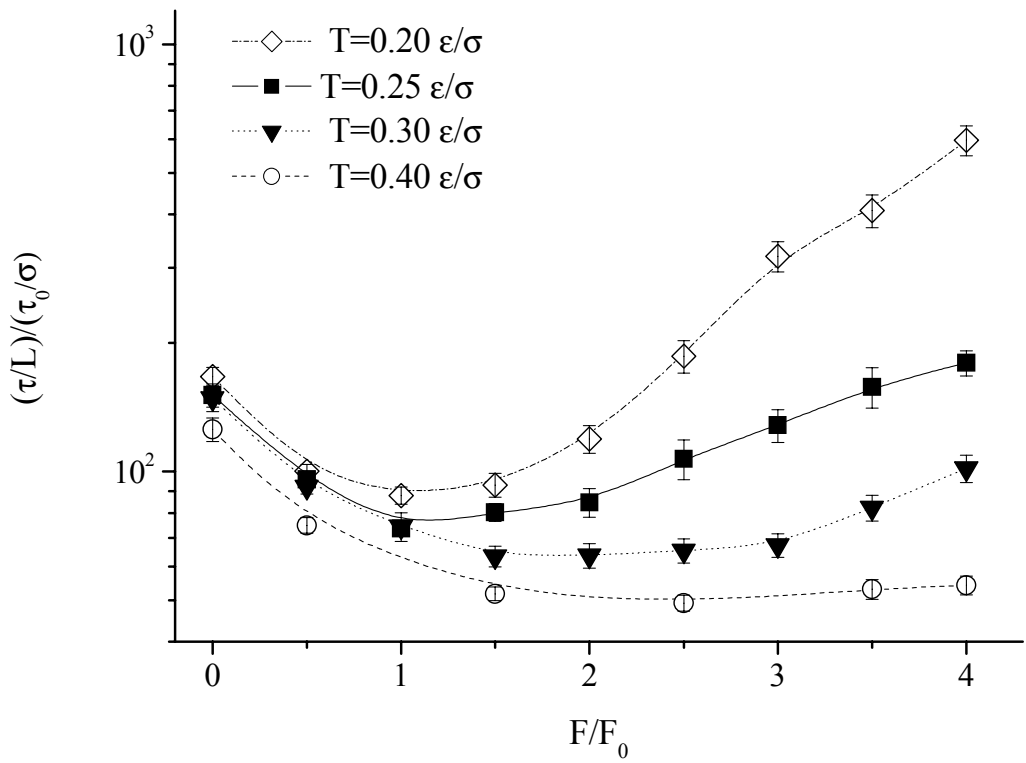

**Figure 4**

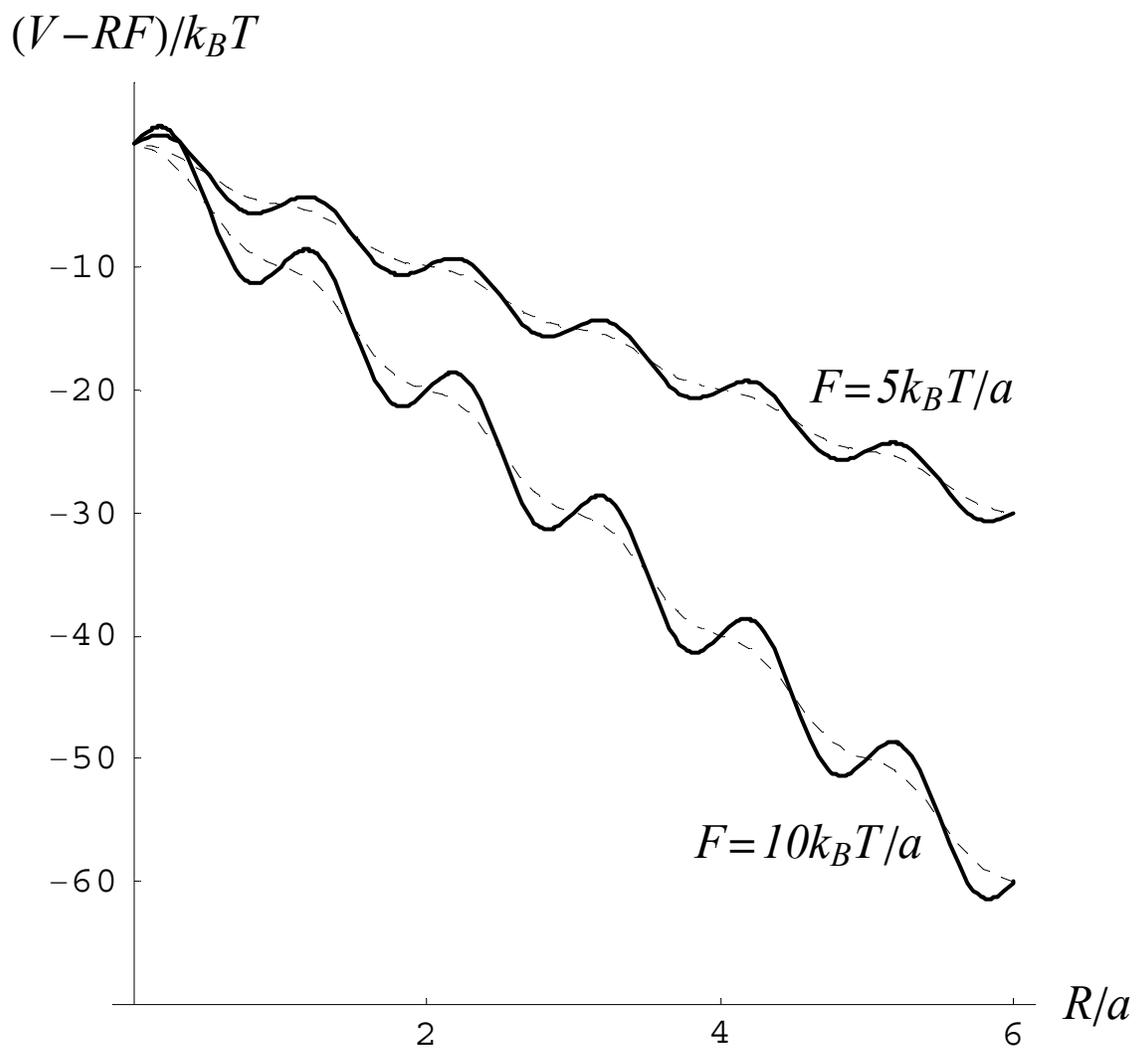

**Figure 5a**

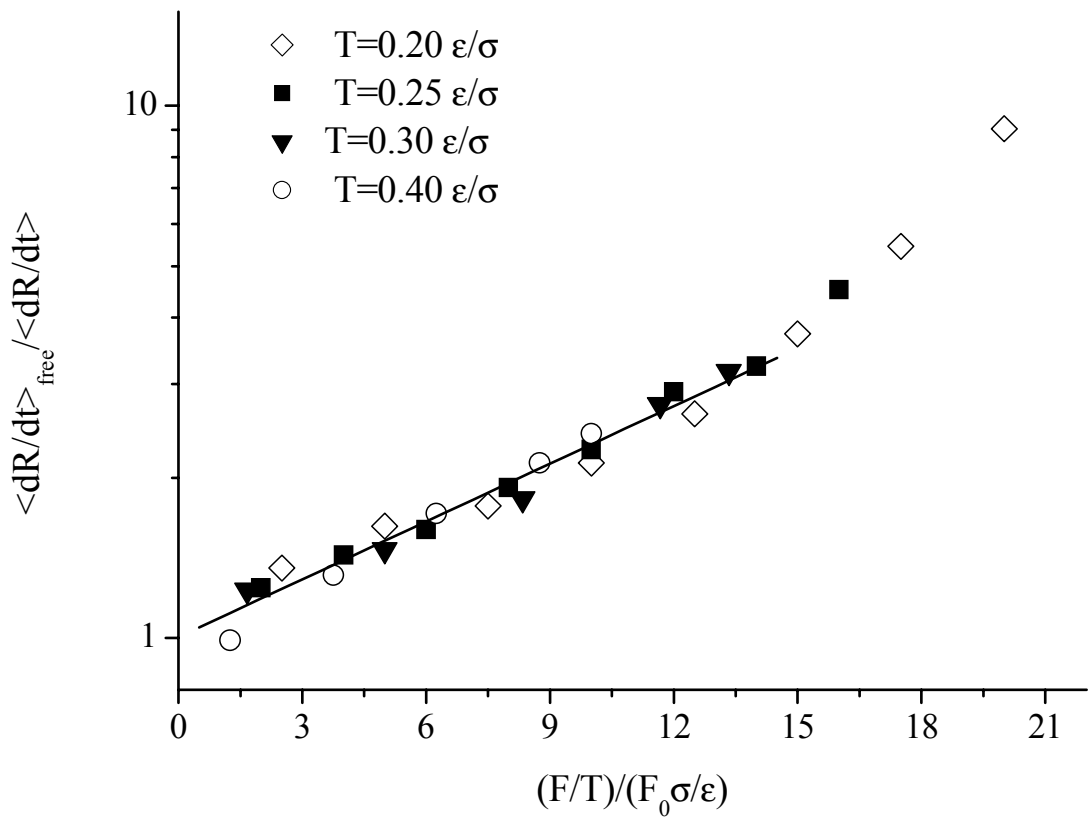

**Figure 5b**